\documentclass[fleqn,usenatbib]{mnras}

\usepackage{newtxtext,newtxmath}

\usepackage[T1]{fontenc}
\usepackage{lineno}

\DeclareRobustCommand{\VAN}[3]{#2}
\let\VANthebibliography\thebibliography
\def\thebibliography{\DeclareRobustCommand{\VAN}[3]{##3}\VANthebibliography}

\usepackage{graphicx}	
\usepackage{amsmath}	

\title[Effects of a Stable Layer on Jupiter’s Jets]{Implications of a Stable Layer on the Vertical Structure of Jet
Streams on Jupiter}

\author[K. Duer-Milner et al.]{
Keren Duer-Milner,$^{1,2}$\thanks{duer@strw.leidenuniv.nl (KDM)}
Louis Siebenaler,$^{1}$
Lillian Haley$^{3}$
and Yamila Miguel$^{1,2}$
\\
$^{1}$Sterrewacht Leiden, Leiden University, The Netherlands\\
$^{2}$SRON - Netherlands Institute of Space Research, Leiden, The Netherlands\\
$^{3}$University of Massachusetts Amherst, Massachusetts, USA
}

\date{Accepted XXX. Received YYY; in original form ZZZ}

\pubyear{\the\year{}}

\begin{document}
\label{firstpage}
\pagerange{\pageref{firstpage}--\pageref{lastpage}}
\maketitle

\begin{abstract}
The vertical structure of Jupiter’s jet streams remains a critical open question for understanding the planet’s atmospheric dynamics and interior. Traditional models often assume an adiabatic density profile, yet recent observations and theory suggest the presence of stable layers, which could significantly alter both the density structure and gravitational signature. We investigate the implications of non-adiabatic stable layers for Jupiter’s gravity field, focusing on how density anomalies from such layers interact with the inferred vertical structure of zonal winds. We construct temperature–pressure profiles including subadiabatic stable layers to derive density profiles consistent with the latest equation of state. The resulting gravitational harmonics are computed, incorporating both static density and wind structure via thermal wind balance, and compared with Juno measurements. By varying the wind decay characteristics, we assess how stable layers constrain the depth and structure of the deep jets. Our results show that shallow, extensive stable layers substantially modify the background density, requiring more rapid decay of zonal winds to satisfy observed gravitational constraints. Introducing stable layers also broadens the range of physically plausible wind solutions, indicating that the vertical structure of the jets is less constrained than suggested by purely adiabatic models. We conclude that stable layers are a critical, yet often overlooked, component in modeling Jupiter’s interior and dynamics. This study highlights a strong degeneracy between the thermodynamic density structure and the vertical wind profile, implying that the jet stream structure cannot be uniquely determined without independent constraints on the planet’s internal stability.
\end{abstract}



\section{Introduction}

The Juno mission has provided high-precision measurements of Jupiter’s gravity field, greatly improving constraints on both its interior structure and its jet streams \citep{Bolton2017,Wahl2017,Iess2018,Kaspi2018}. The gravity field reflects the three-dimensional distribution of density inside the planet, from the deep interior up to the outer atmosphere. Zonal jets modify this background density field, and these density perturbations can leave a detectable imprint on the gravity signal \citep{Hubbard1999}. In particular, Juno’s measurements of odd and high-order gravitational harmonics \citep{Kaspi2013a,kaspi2023} are not expected to be produced by a purely one-dimensional (background) density distribution \citep{hubbard1974}.

Jupiter's immense size and rapid rotation strongly influence its large-scale atmospheric motions, ensuring that the Coriolis force plays a dominant role. Under these conditions, the thermal wind balance emerges as the leading-order vorticity balance of the momentum equation \citep{Pedlosky1987}. This fundamental relationship directly governs the coupling between the planet's persistent zonal flows (the jet streams) and its background radial density structure (hereafter 'static density').  This same principle, linking the large-scale atmospheric dynamics to the underlying thermodynamics variables, is well-established for Earth’s large-scale atmospheric circulation \citep{Vallis2006}.

{The static density profile primarily accounts for the even, low-order gravity harmonics, which probe the deep interior. Higher-degree harmonics are more sensitive to shallower regions because their calculation places progressively greater weight on mass distributions closer to the planetary surface.} Inference of the static density profile from Juno data indicates that Jupiter’s envelope is radially inhomogeneous, with a large dilute core and possibly a small compact core \citep{Wahl2017,Helled2017,miguel2022}. However, many distinct interior density profiles can reproduce the observed even harmonics \citep{militzer2022juno,howard2023jupiter,ziv2024characterizing,helled2024giant}, so the true density structure remains non-unique with current data.

\begin{figure}
   \centering
   \includegraphics[width=\columnwidth]{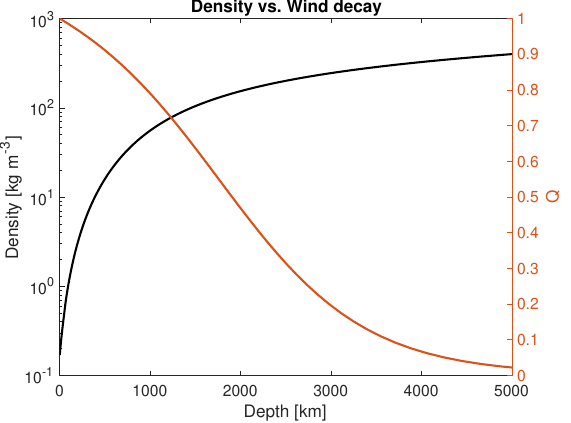}
   \caption{Example of the two vertical components required to match the odd and high-order gravity harmonics: a density profile (here is an adiabat) (black) and a vertical wind decay structure (Q) (orange). The orange profile corresponds to the solution used in \citet{kaspi2023}.}
   \label{fig:rho_vs_Q}
 \end{figure}

Non-uniqueness extends beyond the static density and the planet's interior structure. Many different wind configurations can reproduce the low-order odd gravity harmonics. For example, a random distribution of density anomalies within the planet has been shown to fit the observations well, but this bears little relation to any physical atmospheric process \citep{Kong2018}. Taking the cloud-level winds and assuming they extend inward to about $3000$ km also reproduces the observations \citep{Kaspi2018}; however, the specific shape of the vertical profile remains unconstrained. In particular, an envelope of vertical wind profiles can match the Juno measurements within their uncertainties \citep{duer2020}. All these solutions, nonetheless, require wind penetration of several thousand kilometers, alignment of the flow along cylinders (surfaces of constant angular momentum), and a broadly similar vertical structure (an example vertical structure is shown in orange, Fig.~\ref{fig:rho_vs_Q}. In this profile the winds decay to $20 \%$ of their cloud-level strength at $3000$ km).

Within the framework of fitting the low-order odd gravity harmonics, different studies have emphasized different aspects of the winds’ vertical and meridional structure: \citet{duer2019} showed that including magnetic-field effects, with a rapid decay at depth, still permits a match to the odd harmonics; \citet{galanti2021Con} showed that jets poleward of $\pm 25^\circ$ are not required to explain the gravity signal; and \citet{dietrich2021linking} reached a similar conclusion for jets equatorward of $\pm 21^\circ$. These results were supported by a new method constraining the high-latitude gravity contribution, confirming that most of the gravity signal arises from jets near $\pm 21^\circ-23^\circ$ \citep{kaspi2023}, and by an independent study that did not rely on observed cloud-level winds \citep{cao2023strong}. Additionally, \citet{militzer2022} demonstrated that different jets can penetrate to different depths, and several studies have explored modest variations in cloud-level winds that still allow, and in some cases improve, fits to the gravity data \citep{duer2020,militzer2022,kaspi2020,galanti2021Con,galanti2020}.

Nevertheless, the direct influence of the static density on the jet streams has barely been explored. \citet{militzer2022} matched all harmonics (even and odds) simultaneously by solving for the interior and jet structures together, but because interior density profiles are highly degenerate in the outer, low-density layers, this approach cannot capture the actual coupling between these components. \citet{ziv2024characterizing} performed a similar joint analysis of interior and winds and faces the same limitation.

The static density traditionally used in these models is an adiabatic profile, corresponding to a fully convective interior and atmosphere (black curve in Fig.~\ref{fig:rho_vs_Q}). In studies that consider more complex density structures, such as the two mentioned above, the main challenge remains matching the low even harmonics, so small deviations in the outer layers cannot be fully accounted for. However, the actual density structure is unlikely to be fully adiabatic, and there is growing evidence for stable non-adiabatic layers in Jupiter’s atmosphere and interior. For example, Juno MWR observations indicate a depletion of alkali metals in Jupiter’s atmosphere \citep{bhattacharya2023,aglyamov2025}. This significantly reduces the radiative opacities around the kilobar level and may favor the formation of a subadiabatic stable layer \citep{siebenaler2025conditions}. Additional support for shallow stable layers comes from the low atmospheric CO abundance \citep{Bezard_2002, Bjoraker_2018} and high deep-water abundance measurements \citep{Li_2020}, which are consistent with surpressed vertical mixing in a stable layer \citep{Cavalie_2023}. Such shallow stable layers have been proposed as a mechanism for producing an inverted metallicity gradient in Jupiter’s atmosphere, potentially resolving the tension between the subsolar-to-solar heavy element abundances predicted by interior models and the supersolar abundances measured by the Juno mission and Galileo probe \citep{Howard_2023, Muller_2024}.

Furthermore, deeper stable layers around the megabar level have been the focus of intensive study in recent years. At these pressures, hydrogen and helium become partly immiscible, which can lead to the formation of a stable yet superadiabatic helium rain layer \citep{Stevenson_1977, Markham_2024}. This can separate convective regions in a way that may enable multiple dynamos \citep{wulff2025meaning}, and has been proposed as a key mechanism in explaining Jupiter's magnetic field \citep{moore2022}. Deep stable layers have also been proposed as a mechanism to suppress jet streams in the deep interiors of giant planets \citep{christensen2020,christensen2024quenching,wulff2022} and to promote the formation of multiple jets \citep{gastine2021}.

Given growing evidence for stable layers in Jupiter’s atmosphere and interior, we investigate the direct influence of non-adiabatic interiors, specifically due to subadiabatic stable layers (SL), on Jupiter’s gravity field and the associated vertical wind structure. We present the construction of the density profiles, the thermal-wind balance, and gravitational-harmonic formulations in section \ref{sec:methds}, examine the impact of these profiles on gravity and winds in section \ref{sec:results}, compare our results with other jet-decay profiles from General Circulation Models (GCMs) in section \ref{sec:GCMs} and with Earth atmospheric data in section \ref{sec:ERA5}, and conclude in section \ref{sec:Conclusions}.

 \section{Methods and Model Setup}\label{sec:methds}
 
We explore the implications of a stable layer on Jupiter’s gravity field and jet structure by constructing temperature–pressure (TP) profiles that explicitly include such layers. We then use these TP profiles to compute the corresponding density structure and, together with the zonal winds, reproduce the gravity field. Below, we describe each component of this procedure.

 \subsection{Constructing TP profiles with
stable layers}

We construct TP profiles using the Schwarzschild stability criterion \citep{Schwarzschild_1958}, which compares the local temperature gradient $\nabla_{\rm T}$ with a dry adiabatic gradient $\nabla_{\rm ad}$. According to this criterion, a SL will exist if the local temperature gradient is subadiabatic, i.e. $\nabla_{\rm T} < \nabla_{\rm ad}$. We note that in doing so, we neglect the effects of mean molecular weight gradients and latent heat release due to condensation, both of which can modify the stability criterion \citep{Leconte_2017}; However, these processes are likely negligible within our primary region of interest ($\sim1<p<10^5$ bar), because mean molecular weight gradients should be confined to the dilute core, and substantial latent‑heat release would occur only for water abundances exceeding those of Jupiter \citep{li2020}.

For all TP profiles, we impose a 1-bar boundary condition of $\rm T_{1bar} = 170 \ \rm K$, consistent with Voyager occultation measurements \citep{Gupta_2022}. Additionally, we enforce the presence of a SL within a prescribed pressure interval $[P_{\rm min}, P_{\rm max}]$ by imposing a subadiabatic temperature gradient in this region. Rather than computing the temperature gradient from opacities, we prescribe an artificial subadiabatic gradient that smoothly connects to the surrounding adiabatic regions. For a layer at temperature $T$ and pressure $P$ within the stable region, the temperature gradient is defined as

\begin{equation}
\nabla_T(T,P) = \nabla_{\rm ad}(T,P)
- k \sin^2\left[
\pi \frac{\ln(P/P_{\rm min})}{\ln(P_{\rm max}/P_{\rm min})}
\right], \label{eq:gradT}
\end{equation}

where the amplitude is given by $k = 0.9 \nabla_{\rm ad}(T',P')$. Here $P' = \sqrt{P_{\rm min}P_{\rm max}}$ denotes the midpoint pressure of the SL, and $T'=\textrm{T}_{1\rm bar}(P/1\textrm{bar})^{0.25}$ is the reference temperature used to evaluate $\nabla_{\rm ad}$ at the midpoint. This prescription ensures that the imposed subadiabatic gradient vanishes smoothly at the boundaries of the SL and joins continuously onto the surrounding adiabatic regions.

{The strength of the imposed stable stratification is controlled by the parameter $k$, which determines the departure of the local temperature gradient from the adiabat. Since the Brunt--Väisälä frequency scales as $N^2 \propto (\nabla_{\rm ad}-\nabla_T)$, larger values of $k$ correspond to stronger stable stratification and larger values of $N$. In this study, $k$ is fixed for all stable layers considered, allowing us to isolate the effects of the location and extent of the stable layer. The actual Brunt--Väisälä frequency within Jupiter's interior remains poorly constrained, with estimates varying substantially depending on the assumed origin of the stable stratification and the method used to infer it \citep[e.g.,][]{Leconte_2017,christensen2020,debras2019}, with only one in-situ measurements by the Galileo probe \citep{Seiff1996,magalhaes2002stratification}. Future constraints may become possible through observations of atmospheric dynamics; for example, the depth of Jupiter's polar cyclones inferred from Microwave Radiometer measurements has been proposed as a potential probe of the underlying stratification strength \citep{gavriel2025dynamical}.
}

To study the role of SLs systematically, we consider a representative set of combinations drawn from $P_{\rm min}={1,10,100,1000}$ bar and $P_{\rm max}={10^{3},10^{4},10^{5},10^{6}}$ bar, plus two extreme cases: a very shallow SL ($3-100$ bar) and a very deep SL ($10^{6}-10^{7}$ bar). Because the true physical mechanism producing a SL is uncertain, we intentionally include these not necessarily realistic configurations to explore the full range of possible effects. In total we study 15 SL profiles (in addition to the adiabat), spanning a wide range of plausible locations and extents.

 \begin{figure}
   \centering
   \includegraphics[width=\columnwidth]{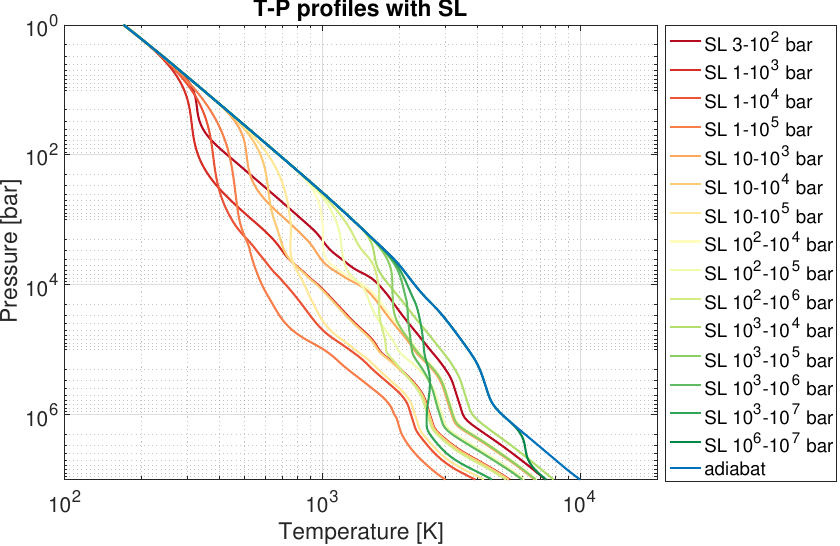}
   \caption{TP profiles that include a stable layer. The depth end extent of the stable layers varies between the profiles as indicated in the legend. Profiles with large stable layers show much colder interiors.}
   \label{fig:TP}
 \end{figure}

  \begin{figure}
   \centering
   \includegraphics[width=\columnwidth]{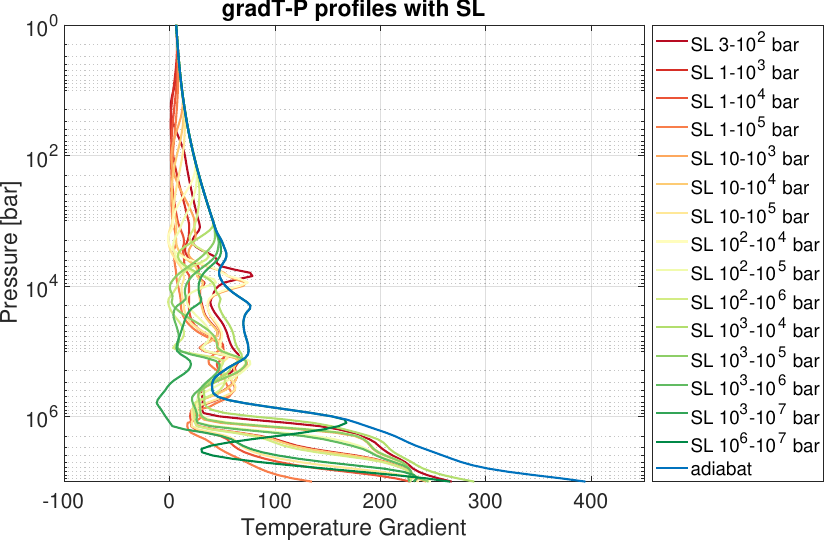}
   \caption{Same as Fig.~\ref{fig:TP} but shown is the temperature gradient with pressure. The stable layers were implemented in the temperature gradient directly as described in Eq.~\ref{eq:gradT}}
   \label{fig:gradTP}
 \end{figure}
 
To compute $\nabla_{\rm ad}$, we use the \citet{Chabrier_2019} EOS for a H-He mixtures and the non-ideal mixing terms from \citet{Howard_2023_NImixing}. Hence,  

\begin{equation}
    \nabla_{\rm ad} =- \frac{(1 - Y) S_{\rm H} \bigg( \frac{\partial \textrm{log} S_{\rm H}}{\partial \textrm{log} P}\bigg)_T + Y S_{\rm He} \bigg( \frac{\partial \textrm{log} S_{\rm He}}{\partial \textrm{log} P}\bigg)_T + \Delta S \bigg( \frac{\partial \textrm{log} \Delta S}{\partial \textrm{log} P}\bigg)_T}{(1 - Y) S_{\rm H} \bigg( \frac{\partial \textrm{log} S_{\rm H}}{\partial \textrm{log} T}\bigg)_P + Y S_{\rm He} \bigg( \frac{\partial \textrm{log} S_{\rm He}}{\partial \textrm{log} T}\bigg)_P + \Delta S \bigg( \frac{\partial \textrm{log} \Delta S}{\partial \textrm{log} T}\bigg)_P},\label{eq:nonidealEOS}
\end{equation}
where $S_{\rm H}$ and $S_{\rm He}$ denote the specific entropy of pure hydrogen and helium, respectively, and $\Delta S = Y (1- Y) S_{\rm mix}$ represents the non-ideal entropy mixing term. We adopt a helium mass fraction $Y=0.238$ throughout the planet, as inferred from in-situ Galileo probe measurements \citep{von_Zahn_1998, Niemann_1998}. We acknowledge that assuming a constant $Y$ and neglecting the contribution of metals is a simplification. However, we found that including $Y$-gradients and a non-zero $Z$ has a negligible impact and does not affect the conclusions of this work. The resulting TP profiles for different SL locations are shown in Fig.~\ref{fig:TP}, with the corresponding temperature gradients shown in Fig.~\ref{fig:gradTP}. 

\begin{figure}
   \centering
   \includegraphics[width=\columnwidth]{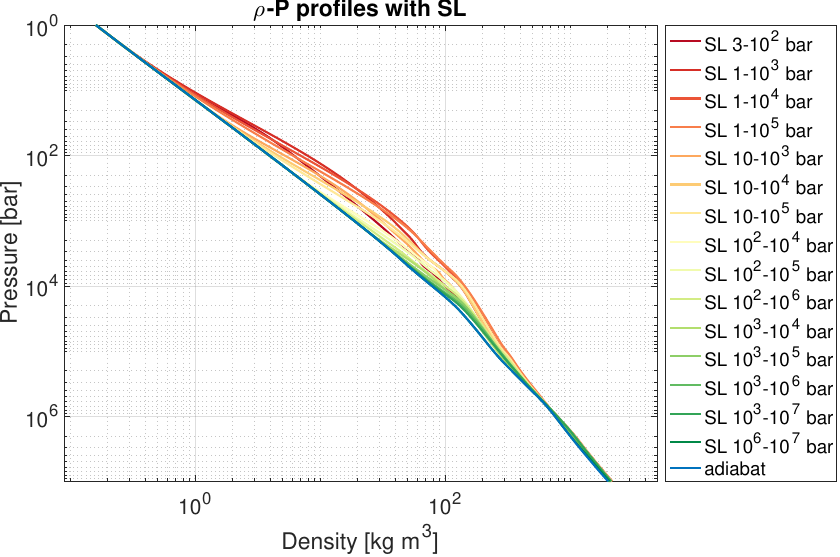}
   \caption{Density-pressure of the 16 profiles from Fig.~\ref{fig:TP}, according to Eq.~\ref{eq:density}. Note that deep stable layers barely influence the density profile.}
   \label{fig:rhoP}
 \end{figure}

\begin{figure}
   \centering
   \includegraphics[width=\columnwidth]{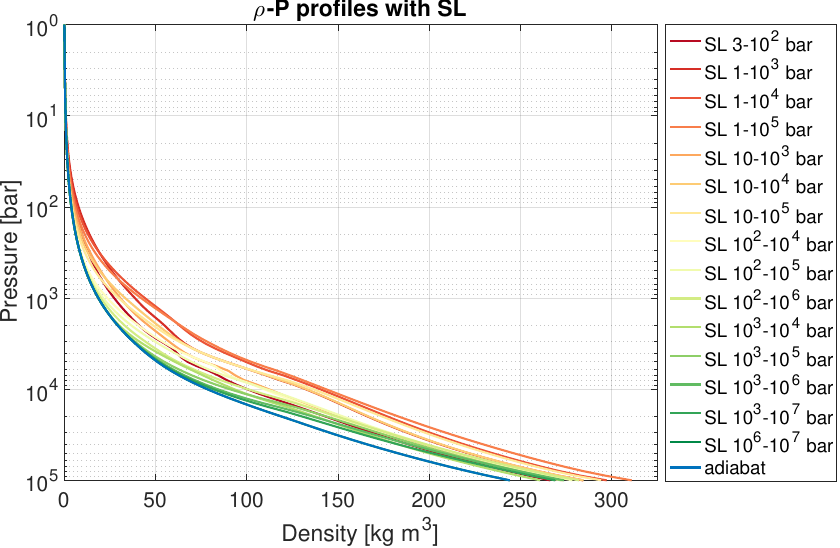}
   \caption{Same as Fig.~\ref{fig:rhoP} (Density-pressure) but in a linear scale for density, focusing on the region where interaction with winds is expected ($p=1-10^5$ bar).}
   \label{fig:rhoP-linear}
 \end{figure}

Mass density profiles are derived using the same EOS, according to

\begin{equation}
    \frac{1}{\rho} = \frac{1-Y}{\rho_{\rm H}} +  \frac{Y}{\rho_{\rm He}} + \Delta V,\label{eq:density}
\end{equation}
where $\rho_{\rm H}$ and $\rho_{\rm He}$ are the densities of pure hydrogen and helium respectively, and $\Delta V = Y(1-Y)V_{\rm mix}$ denotes the non-ideal density mixing term. The resulting density profiles are shown in Figs.~\ref{fig:rhoP} and \ref{fig:rhoP-linear}. SLs can significantly modify the density structure, although their influence is confined to pressures below $\lesssim 1 \ \rm Mbar$. Even though SLs can substantially cool the deep interior, by up to $\Delta T \sim 30000 \ \rm K$ (Fig.~\ref{fig:TP}), the density in these deep layers is unaffected as it is primarily controlled by pressure. At lower pressures, temperature changes due to SLs lead to density variations, with the largest deviations from an adiabatic profile occurring when the stable layer has a shallow upper boundary.

We emphasize that our profiles do not account for a core, as they are only constructed down to a pressure of $10^7$ bar (Fig.~\ref{fig:Jupiter_profiles}). As such, we do not model the region primarily probed by the low-order even gravity harmonics ($J_2$, $J_4$, $J_6$, and $J_8$), which is not the focus of this work. Additionally, we acknowledge that inserting SLs does not strictly conserve mass in our simplified approach; however, the net effect on the planet’s static density is small. The low-order gravity harmonics, which dominate the observable gravity signal, are primarily determined by the mass distribution in the deep interior (see next subsection). As demonstrated above, these deeper regions remain essentially unchanged by the addition of SLs (Fig.~\ref{fig:Jupiter_profiles}). Any mass deficit or excess introduced by the SLs produces a small contribution to the total gravitational moments and could, for example, be compensated by a different core structure. We further do not consider heavy elements, which can also produce a similar effect on the density distribution throughout the planet. Our goal is to isolate the effects of stable layers on the resulting dynamics; additional complexity in the interior structure would likely obscure this effect.


 \subsection{Gravitational harmonics}

The density field of Jupiter can be expressed via the zonal gravity harmonics $\left(J_{n}\right)$, which describe the gravitational field of the planet in equilibrium \citep{zharkov1974}.
The gravity harmonics are

\begin{eqnarray}
J_{n} & = & -\frac{1}{MR_{J}^{n}}\int\rho r^{n}P_{n}d^{3}r,\label{eq:Jn}
\end{eqnarray}
where $M$ is Jupiter's mass, $R_{J}$ is Jupiter's equatorial radius, $n$ is the harmonic degree {($n=2,...,N$)},
$\rho$ is the density, $r$ is the radial coordinate and $P_{n}$ is the $n$-th Legendre polynomial \citep{Hubbard1984}. The density can be decomposed into a static component, representing the planetary interior structure \citep[e.g., ][]{miguel2022}, and a dynamical component, corresponding to density anomaly resulting from redistribution of mass by fluid velocities (with respect to interior rotation) \citep{Kaspi2010a}. This gives:

\begin{eqnarray}
\rho\left(r,\theta\right)=\rho_{\rm{stat}}\left(r\right)+\rho'\left(r,\theta\right),
\end{eqnarray}
where $\theta$
is latitude. The zonal gravity harmonics corresponding to the dynamical part of the flow $\left(\Delta J_{n}\right)$
can be calculated by integrating the density anomaly and its projection
onto the Legendre polynomials in spherical coordinates such that

\begin{eqnarray}
\Delta J_{n} & = & -\frac{2\pi}{MR_{J}^{n}}\intop_{0}^{R_{J}}\intop_{-1}^{1}\rho^{\prime}\left(r,\mu\right)r^{n+2}P_{n}\left(\mu\right)d\mu dr,\label{eq:anom_Jn}
\end{eqnarray}
where $\mu=r\sin\theta$. {A planet with no dynamics is symmetric between north and south, and therefore has no static contribution to the odd gravity harmonics. Consequently, for odd harmonics, the measured gravity moments arise entirely from the dynamical density anomaly, such that $\Delta J_n = J_n$ for $n=3,5,\ldots$.} The density anomaly represented by the odd harmonics should therefore be negligible if the flow pattern is symmetric or if the dynamics are very shallow. Furthermore, the high gravity harmonics by a rotating planet with no winds will be very small \citep[e.g.,][]{nettelmann2021}. Yet, for Jupiter, Juno measured both odd and the high-order gravity harmonics,
which can only be attributed to $\rho'$ \citep{Iess2018,kaspi2023,sun2026estimation}.

Using the SL density profiles derived above and the wind contribution described below, we compute the dynamical density anomalies and the resulting dynamical gravity harmonics (the low-order odd harmonics and the high-order harmonics from $J_{10}$ to $J_{20}$), and compare the modeled harmonics with the Juno observations reported by \citet{kaspi2023}. We omit the low‑order even harmonics ($J_4,J_6,J_8$), since these are strongly dependent on the assumed interior density profile.

 \subsection{Thermal-wind Balance}

The relationship between the flow field and the dynamical density can be expressed via the thermal wind balance \citep[e.g., ][]{Pedlosky1987,Holton1992}. This balance is the leading order vorticity balance for planets with small Rossby numbers, which is the case of Jupiter, as well as all the other giant planets in the Solar System \citep{duer2025gas}. If only zonal jets are considered, which is valid for Jupiter as the other flow components are expected to be at least an order of magnitude smaller \citep{duer2021}, the thermal wind balance
becomes

 \begin{equation}
   2\Omega r \frac{\partial (U \rho_{\rm{stat}})}{\partial z} = g_0 \frac{\partial \rho'}{\partial \theta}.
   \label{eq:thermal_wind}
 \end{equation}

 where \( \Omega \) is the rotation rate, \( U \) is the zonal wind speed, \( z \) is the vertical coordinate, and \( g_0 \) is the gravitational acceleration. An equivalent equation can be written with temperature instead of density gradients, and one can easily switch between the two versions \citep{Kaspi2016}.  Higher order
expansion (dynamic self-gravity), beyond thermal wind, provides a relative small
correction for determining the deep flow dynamics, hence, we neglect these here \citep{Galanti2017a,Galanti2017b,Zhang2015,wicht2020linking}.

The deep wind structure is composed by multiplying the cloud-level winds with a radial decay function $Q$. An example for such a decay profile is available at Fig.~\ref{fig:rho_vs_Q}. This gives:

 \begin{equation}
   U = Q U_{\rm{cyl}}.
   \label{eq:wind}
 \end{equation}
where $U_{\rm{cyl}}$ is the cloud-level winds projected in the cylindrical (vertical) direction inwards. We can then expand the right-hand-side of Eq. \ref{eq:thermal_wind} to understand the interaction between the decay function and the static density:

\begin{equation}
2\Omega r \frac{\partial (Q U_{\rm{cyl}} \rho_{\rm{stat}})}{\partial z} = 2\Omega U_{\rm{cyl}} r (\rho_{\rm{stat}}\frac{\partial Q}{\partial z} + Q \frac{\partial \rho_{\rm{stat}}}{\partial z} ).
\label{eq:TW_exp}
\end{equation}

Note that the two terms in the right-hand-side of Eq. \ref{eq:TW_exp} have a similar  contribution to the overall structure of the dynamical density. 

 \subsection{Vertical wind structure}

The vertical structure of the zonal flow is poorly constrained, and currently, no robust physical constraints are required when deriving a profile to match the gravity harmonics. The only direct measurement comes from the Galileo probe in 1995, which measured the flow at $7^\circ$ N. This revealed an increase in wind speed from $\sim 80$ ms$^{-1}$ at cloud level to $\sim 160$ ms$^{-1}$ at 4 bar, followed by constant wind strength down to 20 bar, where transmission failed \citep{Atkinson1996,atkinson1998galileo}. Although previous studies showed that incorporating the Galileo trend allows for fitting gravity measurements with a different vertical structure \citep{Kaspi2018,galanti2021Con}, the Galileo entry site was identified as a localized hot spot \citep{Showman2000} and is therefore typically excluded when fitting global trends. {Furthermore, as we demonstrate below, fitting the gravity data is sensitive only to the deeper regions of the atmosphere, where the density is higher. Hence, changes to the winds near the cloud layer have only a very small effect on the inferred deep wind solution.} At greater depths ($\sim 0.93 R_J$) the flow strength is expected to be negligible. Studies investigating the interaction between the magnetic field and the winds indicate velocities of order centimeters per second in these deep regions \citep{duer2019,moore2019,bloxham2022differential,bloxham2024rapidly}. Accordingly, here we consider only a decay in flow strength from its peak at cloud level to these negligible values in the interior.

We solve Eq.~\ref{eq:thermal_wind} by projecting the observed cloud‑level winds inward along cylinders (no modification of the latitudinal structure) and computing the resulting dynamical density anomalies for each SL density profile. For each profile we either (i) adopt a pre‑specified decay function $Q$ and compute the harmonics directly via Eq.\ref{eq:anom_Jn}, or (ii) determine the best‑fit vertical decay by optimizing $Q$ using one of the two approaches described below (constrained parametric or a free‑decay). The optimization procedure is summarized in Appendix~\ref{app:optim}.

 \subsubsection{Constrained vertical wind structure}

The vertical structure of the zonal flow can be defined by three independent parameters that together can represent a wide range of vertical structures \citep{Kaspi2018,duer2020}.
It is set as

\begin{equation}
Q(r)=
\begin{array}{cc}
(1-\alpha)\exp\left(\frac{r-R_{J}}{H}\right)+\alpha\left[\frac{{\rm tanh}\left(-\frac{R_{J}-H-r}{\Delta H}\right)+1}{{\rm tanh}\left(\frac{H}{\Delta H}\right)+1}\right],
\end{array}\label{eq:decay}
\end{equation}
where $\alpha$ defines the ratio between the exponential decay and the hyperbolic tangent components, $H$ represents the decay scale height (or ‘half-life’), and $\Delta H$ is the width of the hyperbolic tangent transition. We treat these three variables as free parameters when solving for the best-fitting solution that will match the observations, as described in Appendix~\ref{app:optim}.

 \subsubsection{Freely decaying}

Another approach is to directly determine the wind values at each grid point. While this method substantially increases the number of free parameters, it removes the potentially non-physical constraints imposed by assuming a specific decay function. To ensure the resulting profile is physically plausible (e.g., decaying or constant with depth), we enforce that

 \begin{equation}
A = \begin{bmatrix} 
1 & -1 & 0 & 0 & \cdots & 0 & 0 \\
0 & 1 & -1 & 0 & \cdots & 0 & 0 \\
0 & 0 & 1 & -1 & \cdots & 0 & 0 \\
0 & 0 & 0 & 1 & \cdots & 0 & 0 \\
\vdots & \vdots & \vdots & \vdots & \ddots & \vdots & \vdots \\
0 & 0 & 0 & 0 & \cdots & 1 & -1 \\
0 & 0 & 0 & 0 & \cdots & & 1
\end{bmatrix}_{n \times n}
 \end{equation}
where $n$ is the number of grid points in the radial direction, and

  \begin{equation}
A\cdot x \leq b,\label{eq:freedecay}
\end{equation}
where, $x$ represents the values of $Q$ to be optimized, and $b$ is a vector of zeros. We set the flow strength to $1$ at the cloud level and to a negligible value $(0.01)$ at $0.94R_{\rm J}$ and deeper. This boundary condition models the effect of an unknown mechanism (e.g., an Ekman layer, semi-conductive layer, or stable layer) that could induce a rapid decay in the winds. This is consistent with studies considering magnetic field effects on the flow field \citep{cao2017,duer2019,moore2019,bloxham2022differential,bloxham2024rapidly}. The remaining $n=120$ grid points are then optimized to find the best-fitting solution.

 \section{Results}\label{sec:results}

 \subsection{The direct effect of stable layers}
\begin{figure*}
   \centering
   \includegraphics[width=0.95\textwidth]{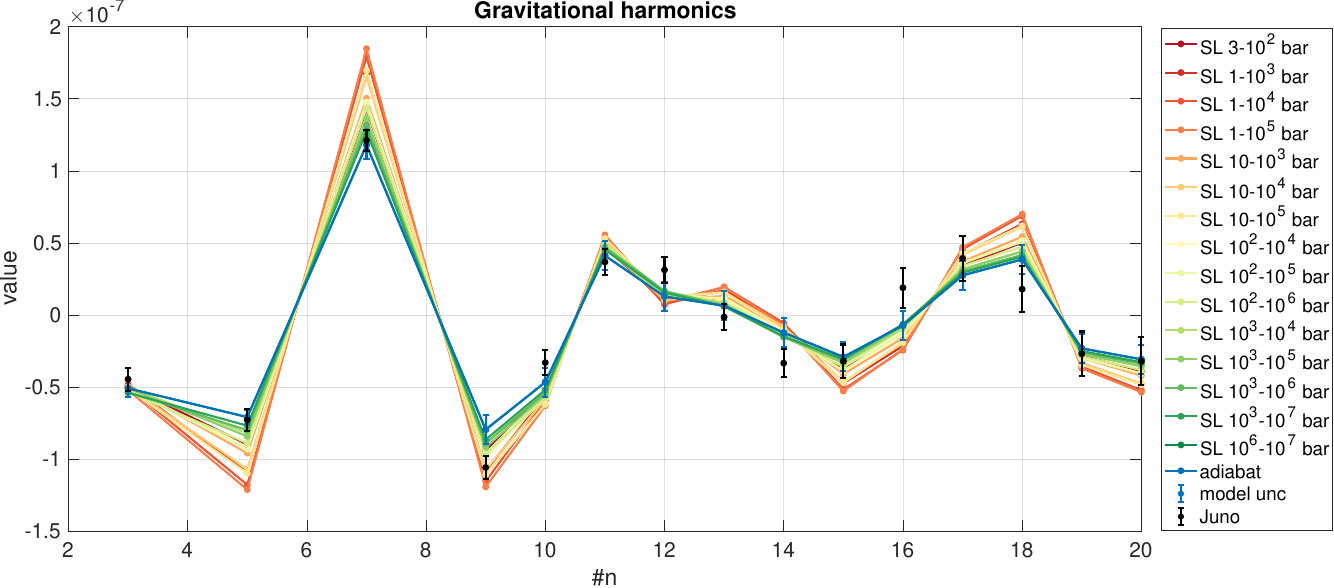}
   \caption{Gravitational harmonics associated with different densities that include a SL, and a constant decay profile. The densities are the 16 profiles shown in Fig.~\ref{fig:rhoP}, and the decay profile is the one shown in Fig.~\ref{fig:rho_vs_Q} (orange). Note we omit the low order even harmonics $(J_4,J_6,J_8)$ as these strongly dependent on the full interior density profile. The error bar is shown only for the solution for the adiabatic profile. Juno's inferred gravity harmonics are shown in black along with $3\sigma$.}
   \label{fig:moments}
\end{figure*}

We introduce various SLs into the static density profile ($\rho_{\rm{stat}}$) and calculate the resulting gravitational harmonics. For these calculations, we use a constant $Q$ profile, specifically the orange profile shown in Fig.~\ref{fig:rho_vs_Q}. This particular $Q$ profile was previously used to explain the higher-order gravitational harmonics \citep{kaspi2023}, and it can be generated using Eq. \ref{eq:decay} with parameters $H=2101 \; \rm{km}$, $\Delta H = 842 \; \rm{km}$, and $\alpha=0.68$. In total, we analyze 16 density profiles, including a purely adiabatic profile calculated under identical conditions and using the same EOS. Model uncertainty curves are presented solely for the adiabatic profile, and are coming from uncertainty in the cloud-level wind speed measurements. Juno's measurements and their uncertainties are taken from \citet{kaspi2023}, and are also displayed in Fig.~\ref{fig:moments}. We show the low-order odd gravity harmonics and the high-order gravity harmonics from $J_{10}$ to $J_{20}$. While this trend extends up to $J_{40}$, higher harmonics are omitted for clarity.

While $J_3$ is largely unaffected by the presence of these SLs, the other harmonics clearly deviate from the adiabatic values. This aligns with findings that $J_3$ is sensitive to deeper regions \citep{duer2020}, where the inclusion of stable layers does not significantly alter the density profile (Figs.~\ref{fig:rhoP},\ref{fig:rhoP-linear}). The observed changes relative to the adiabatic case are primarily influenced by two factors: the stable layer's position and its extent. Shallower and more extensive SLs cause the harmonic values to diverge more strongly. Conversely, deep SLs, although located in regions of high density, have minimal impact on the overall density structure. This is because at such high pressures, temperature variations negligibly affect density as demonstrated in Fig.~\ref{fig:rhoP}. Therefore, deep SLs are not expected to significantly influence the gravity field.

We note that the effect of shallow and extensive SLs is equivalent to that of an adiabatic density profile combined with deeper winds (see Fig.3 and ED Fig.2 in \citet{kaspi2023}). While the impact of deeper winds is relatively intuitive, it is similar to increasing the amplitude of a pulse, which results in a greater wave amplitude in the spectral domain, achieving the same effect through stable layers is less immediately apparent. This can be explained by simply examining Eq.~\ref{eq:TW_exp}: instead of increasing $\frac{\partial Q}{\partial z}$ (due to stronger winds), we are effectively increasing $\frac{\partial \rho_{\rm{stat}}}{\partial z}$ (due to SLs). Both mechanisms lead to a similar trend in the gravitational harmonics. Consequently, these shallow, extensive SLs, when combined with the chosen vertical jet structure, no longer provide a good fit to the observed signal.
 
 \subsection{Constrained optimal decay solution with SL}

To determine the optimal vertical jet structure corresponding to density profiles with SLs, we search for the best-fit parameters of the constrained decay function (Eq.~\ref{eq:decay}) that minimize the cost function (Eq.~\ref{eq:cost}). For this optimization, we primarily consider the low-order odd gravity harmonics ($J_3, J_5, J_7,$ and $J_9$). However, the resulting trends and similar values are consistently observed for the high-order gravity harmonics as well.

    \begin{figure}
   \centering
   \includegraphics[width=1.01\columnwidth]{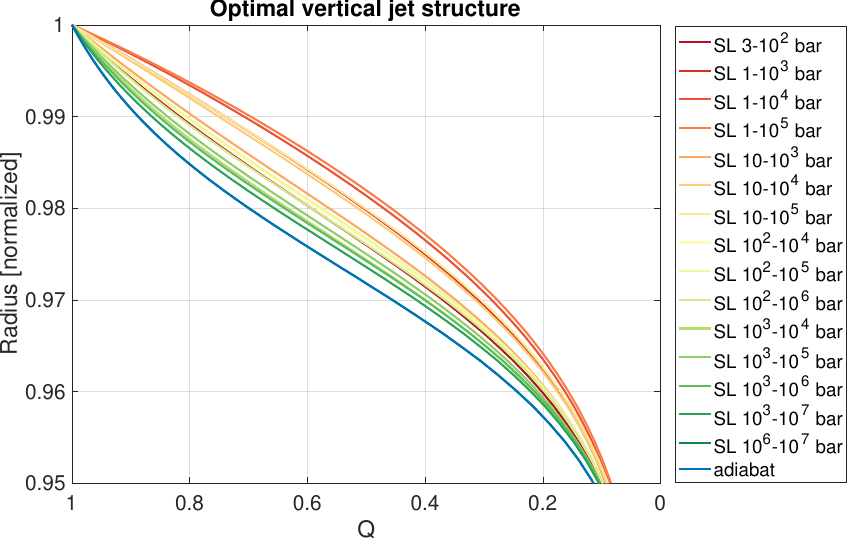}
   \caption{Constrained optimal vertical wind decay structures for profiles with a SL, shown for the outer $5\%$ of Jupiter's radius. These profiles were achieved by fitting the low order odd gravity harmonics $(J_3,J_5,J_7,J_9)$, by finding 3 free parameters defined in Eq.~\ref{eq:decay}. Shallow and extended stable layers result in weaker winds at the same depth relative to the adiabat.}
   \label{fig:wind_decay}
 \end{figure}

The resultant optimal vertical jet structures reveal the intricate interplay between the vertical shape of $\rho_{\rm{stat}}$ and $Q$. Density profiles that incorporate a SL exhibit larger densities at pressures within the SL, and up to $\sim10^6$ bar, compared to an adiabatic profile (Figs.~\ref{fig:rhoP},\ref{fig:rhoP-linear}). Consequently, the winds must be weaker to compensate and maintain the observed gravity field (Fig.~\ref{fig:wind_decay}). Thus, profiles with SLs necessitate faster wind decay to produce the same total density perturbation, which contributes to the gravitational harmonics. Given that the existence of SLs in Jupiter's upper atmosphere is not yet confirmed, we cannot definitively determine the planet's "true" vertical jet structure. However, the interdependence of these terms implies that constraining one will provide valuable insights into the other.

 \subsection{Free optimal decay solution}

Next, we relax the constraint on the vertical structure and explore solutions obtained using a freely decaying wind profile (Eq.\ref{eq:freedecay}). We examine two scenarios: fitting solely the low-order odd gravity harmonics, and simultaneously fitting both low-order and high-order gravity harmonics. These analyses are performed for both an adiabatic density profile and for the most shallow and extensive SL profile, located between $1$ and $10^5$ bar. As demonstrated in the previous experiment (Fig.\ref{fig:wind_decay}), other less extensive or deeper SLs are expected to yield results intermediate to these two extremes.

\begin{figure}
   \centering
   \includegraphics[width=\columnwidth]{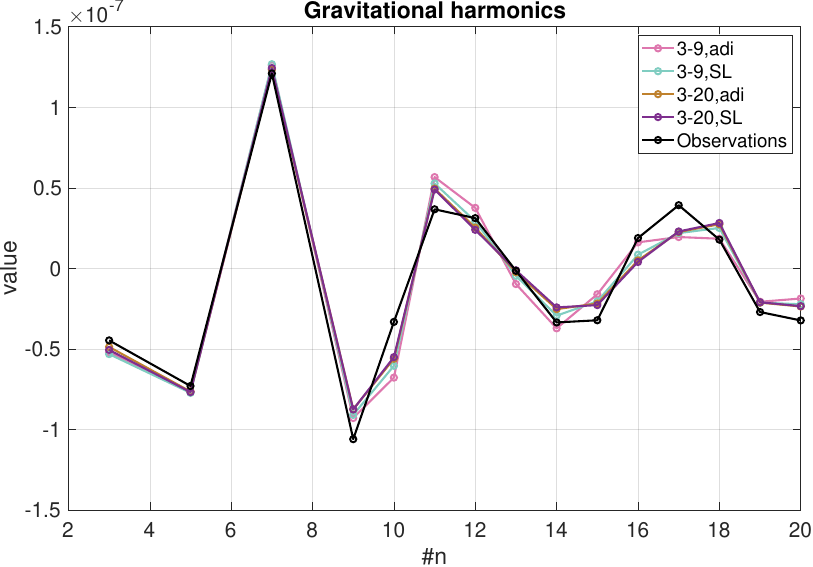}
   \caption{Gravitational harmonics obtained from the optimal solutions using a free decay profile (Eq.~\ref{eq:freedecay}). Shown are the solutions for an adiabatic profile and for the profile containing the most extensive stable layer ($1$–$10^5$ bar), obtained by fitting either the odd harmonics ($J_3$, $J_5$, $J_7$, and $J_9$) alone or together with the high-order harmonics.}
   \label{fig:moments_free_decay}
 \end{figure}

Evidently, relaxing the constraints on the vertical profiles still allows for finding excellent solutions for the gravity field (Fig.~\ref{fig:moments_free_decay}). Interestingly, the poorest fit is observed for the adiabatic profile when only the low-order odd harmonics are fitted (pink curve), which represents an approach commonly taken (though typically with a constrained decay function). Nevertheless, given the high degeneracy of this problem, all tested configurations provide highly satisfactory solutions.

\begin{figure}
   \centering
   \includegraphics[width=1\columnwidth]{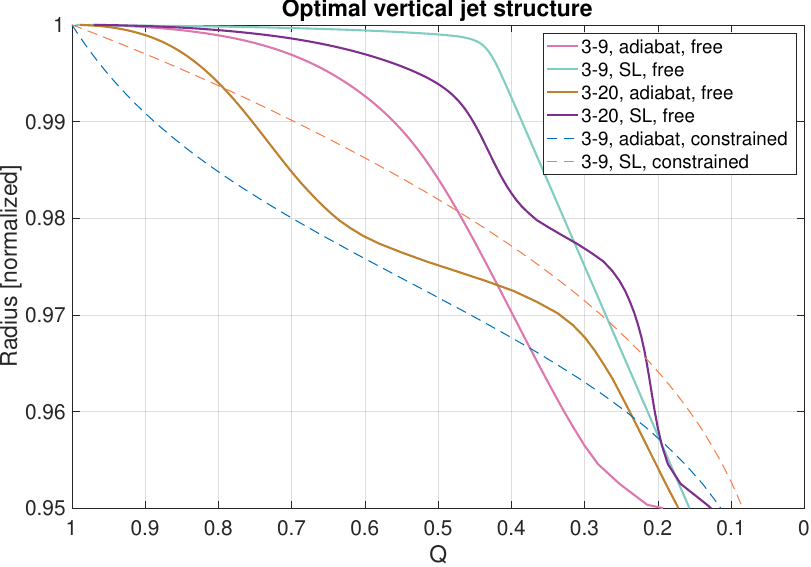}
   \caption{Freely decaying vertical wind structures (full lines), corresponding to the harmonics shown in Fig.~\ref{fig:moments_free_decay}, and the two extreme constrained vertical wind structures (dashed) from Fig.~\ref{fig:wind_decay}. The brown and purple lines were obtained by fitting the odd and the high gravity harmonics simultaneously.}
   \label{fig:wind_decay_free}
 \end{figure}

We further compare the vertical jet structure to the results of the constrained experiment (Fig.\ref{fig:wind_decay_free}). The two extreme cases from the constrained experiment are shown as dashed lines, while the freely-decaying profiles are shown as solid lines. The trend revealed in the constrained experiment (Fig.\ref{fig:wind_decay}) becomes even more pronounced: shallow and extensive SLs necessitate significantly weaker flow strengths within the SL's region. In fact, the freely-decaying solution for an extensive SL (turquoise line) demonstrates winds decaying to less than half of the cloud-level strength within the uppermost $0.2\%$ of Jupiter's radius (approximately the top $100$ km).

Simultaneously fitting the odd and high-order gravity harmonics reveals a different trend than fitting only the low-order odd harmonics. The high-order harmonics appear to require stronger winds closer to the cloud level, effectively resulting in vertical profiles that exhibit two distinct regions of rapid decay: one near the clouds and a second between $0.97$ and $0.98$ $R_{\rm{J}}$. While this feature may not directly reflect a simple physical mechanism, it is nonetheless significant because forcing the high-order harmonics to match observations as closely as possible necessitates stronger winds in shallower regions. This likely reflects the greater sensitivity of high-order harmonics to weaker, higher-latitude jet streams that rapidly become negligible at larger depths.

{The opposite behavior is expected for super-adiabatic layers. Rather than increasing the density relative to the adiabatic profile, these layers produce localized density deficits. To reproduce the same gravitational signature, this reduced density contribution must be compensated by a slower decay of the zonal winds with depth. As a result, stronger winds persist into the deeper, denser regions of the planet, where they contribute more significantly to the gravity field and compensate for the reduced contribution from the super-adiabatic layer.
}

Overall, the inclusion of stable layers{, or more generally, non-adiabatic layers,} demonstrates that the range of possible vertical wind structures is considerably broader than previously assumed. This introduces an additional degree of freedom that must be taken into account when interpreting Jupiter's gravity field and constraining its atmospheric dynamics and interior structure.

\subsection{Exploring Jet Decay within the Stable Layer}

{We further considered a physically motivated flow structure in which the jets decay within the stable layer. Theoretical and numerical studies \citep{christensen2020,christensen2024quenching} suggest that convectively driven zonal jets are attenuated within stably stratified regions. These studies find that the penetration depth depends on both the stratification strength and the characteristic jet width, with stronger stratification leading to shallower penetration while still allowing the jets to extend below the upper boundary of the stable layer. The prominent jets on Jupiter are separated by several degrees in latitude, corresponding to characteristic widths of a few thousand kilometers. Since the stratification strength is not varied in this work, we adopt a simplified prescription in which the jets decay approximately $1000$\,km below the upper boundary of the stable layer. Specifically, we use Eq.~\ref{eq:decay} with $\alpha=1$, $\Delta H=1000$\,km, and set $H$ equal to the depth of the upper boundary of the stable layer.}

\begin{figure*}
   \centering
   \includegraphics[width=0.95\textwidth]{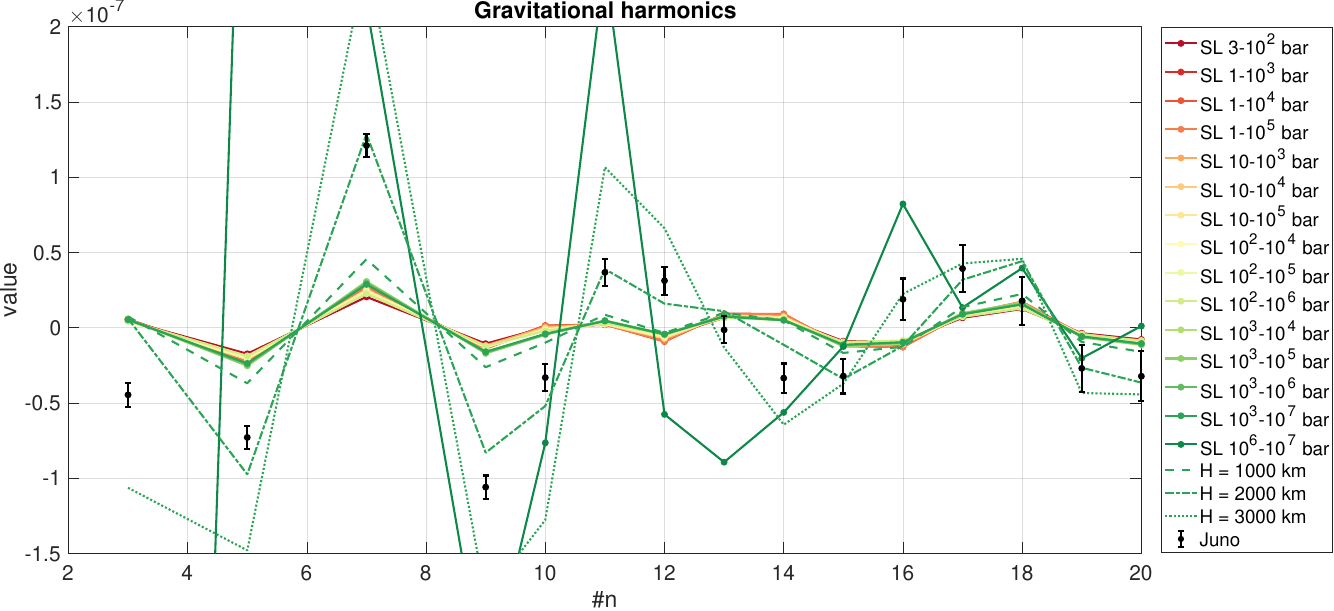}
   \caption{Gravitational harmonics resulting from the different SLs in which the zonal winds begin to decay at the upper boundary of the SL and attenuate over a characteristic depth of $1000$\,km. For the extended SL spanning $10^3$--$10^7$\,bar, we additionally consider decay profiles that begin at depths of $1000$, $2000$, and $3000$\,km below the cloud tops, corresponding approximately to pressures of $5\times10^3$, $3.5\times10^4$, and $8\times10^4$\,bar, respectively. The gravity harmonics inferred from Juno are shown in black, with the corresponding $3\sigma$ uncertainties.}
   \label{fig:moments_rev}
\end{figure*}

{As shown in Fig.~\ref{fig:moments_rev}, forcing the jet decay to occur within the stable layer does not allow the Juno gravity data to be reproduced. All 14 shallow stable-layer models yield gravity harmonics that are too weak, as the jets decay before reaching sufficient depths and enclosed mass to generate the observed signal. Conversely, the deepest stable-layer model considered ($10^6$--$10^7$\,bar) produces gravity harmonics that are too large in amplitude and exhibit an incorrect relative structure. In this case, the jets penetrate too deeply, and the contribution from the high-latitude jets becomes excessively strong, shifting the moments away from the observed ratios.}

{We note that fitting the Juno gravity data requires the zonal winds to retain at least $\sim10\%$ of their cloud-level amplitude at depths of approximately $3000$\,km below the cloud tops ($\sim0.95\,R_J$, corresponding to pressures of order $10^5$\,bar; see Fig.~\ref{fig:hydro}). Therefore, any prescription that forces the jets to decay substantially above this depth is unable to reproduce the observed gravity harmonics.}

{For one of the extended stable-layer models ($10^3$--$10^7$\,bar), we further considered the possibility that the decay does not begin at the upper boundary of the stable layer but deeper within it. Specifically, we tested decay profiles with $H=1000$, $2000$, and $3000$\,km below the cloud tops, corresponding approximately to pressures of $5000$, $35000$, and $80000$\,bar, respectively. These depths are roughly representative of the middle of the stable layer, although they are not directly tied to any specific physical transition within it. This experiment demonstrates that the simplified decay prescription can still reproduce the Juno gravity harmonics reasonably well, provided that the winds retain sufficient amplitude down to the depth constraint discussed above.}

  \section{Vertical jet structure from Jupiter GCMs}\label{sec:GCMs}

As previously established, the gravity-inversion problem is non-unique and the inclusion of density-profile effects further exacerbates this degeneracy. To compare the resulting envelope of possible vertical flow structures with physically driven constraints, we analyze the vertical structure of jet streams from two distinct Jupiter General Circulation Models (GCMs), both of which simulate eddy-driven flow, and compare them with solutions obtained from gravity inversion.

We specifically focus on eddy-driven jets, which are characteristic of Jupiter's higher latitudes \citep{Salyk2006,Choi2011,galperin2014cassini,young2017,duer2021}. The equatorial jet, on the other hand, is unlikely to be associated with eddy momentum flux convergence in the meridional direction \citep{duer2024}, nor does it significantly influence the gravity field due to the cylindrical projection of the winds \citep{duer2020,kaspi2023,Fletcher2021}, leading to its effective vanishing near the cloud level.

{The first GCM explored is a "deep" model driven by convecting plumes, based on results from direct numerical simulations (DNS) performed with the Rayleigh convection code \citep{featherstone2016,Featherstone2022,duer2023}. To isolate the effects of deep convection, the simulations assume a fully convective shell and do not include stable stratified layers, radial variations in electrical conductivity, or magnetic (Lorentz-force) effects. Consequently, no mechanism is present to attenuate the zonal winds with depth, allowing the resulting jets to persist throughout the computational domain.}

\begin{figure}
   \centering
   \includegraphics[width=\columnwidth]{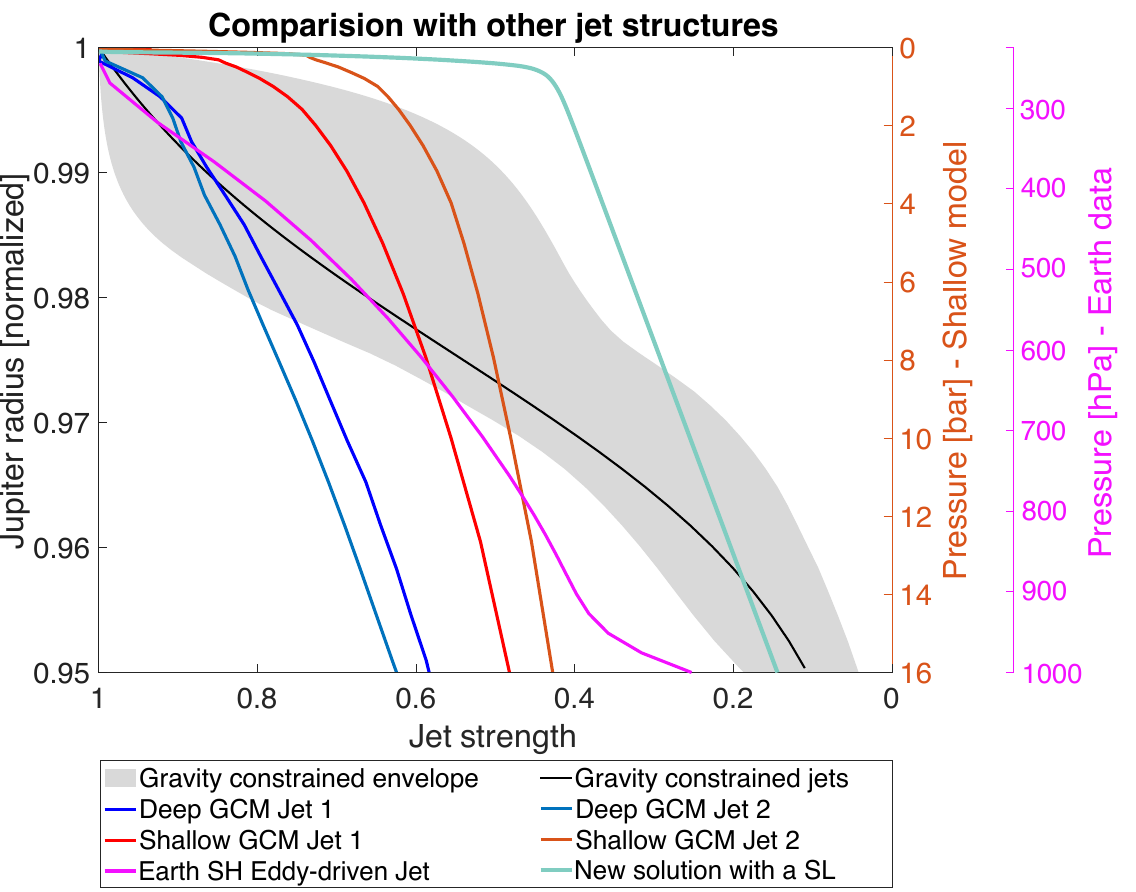}
   \caption{Comparison of vertical jet‑decay structures from gravity inversions, GCMs, and Earth data. Shown are the constrained best‑fit gravity solution for the adiabat (black; same as in Fig.~\ref{fig:wind_decay}), a candidate solution with a shallow SL (turquoise; same as in Fig.~\ref{fig:wind_decay_free}), vertical profiles from a deep Jupiter GCM \citep[blue lines;][]{duer2023}, a shallow Jupiter GCM \citep[red/orange lines;][]{young2019}, and the climatological eddy‑driven jet on Earth (pink). Each profile is normalized by its local maximum and plotted against its native vertical coordinate; accordingly different curves use different y‑axes (blue and black: left axis; orange/red: right (orange) axis; pink: rightmost (pink) axis). The gray shaded region shows the $9\sigma$ envelope of decay structures obtained by fitting the low‑order gravity harmonics with an adiabatic density profile using a constrained decay form \citep[from][]{duer2020}. This comparison highlights the wide range of physically plausible decay behaviors and how SLs expand the envelope of solutions derived from gravity inversions.}
   \label{fig:GCMs}
 \end{figure}

We compare the vertical structure of these jets in the upper $5\%$ of Jupiter's radius (Fig.~\ref{fig:GCMs}) to solutions from other GCMs and gravity inversion results. We track two distinct jet maxima in the southern hemisphere, noting that this particular simulation exhibits near-perfect north-south symmetry. Since this simulation does not incorporate a frictional mechanism for the winds, the decay is slow, and the jets maintain their strength relative to solutions derived from gravity inversion. Nevertheless, the jets do exhibit a relatively linear decay trend. All jet values are normalized by their maximum to facilitate proper comparison across different models and with observations. The full wind velocity field for this model is provided in the appendix (Appendix~\ref{app:GCM_duer}, Figs.~\ref{fig:duerGCM1},~\ref{fig:duerGCM2}).

The second GCM is a 'shallow' model from \citet{young2019}, implemented using the MITgcm \citep{Adcroft2007}. This model simulates Jupiter's weather layer, extending between 1 and 18 bars. A comprehensive analysis of its force balance, revealing the dominance of eddy terms in mid-latitudes, is presented in \citet{young2019}. We again track two distinct jet maxima in the southern hemisphere. The full wind velocity field for this model is provided in the appendix (Appendix~\ref{app:GCM_young}, Fig.~\ref{fig:youngGCM}). These jets exhibit a rapid exponential decay: the wind strength drops quickly near the cloud level and then remains relatively constant until the bottom of the simulation domain.

 \section{Analysis of ERA5 Climatology Data for Eddy Driven Jets}\label{sec:ERA5}

Lastly, to further our understanding of eddy-driven jets, we utilize the ERA5 climatology data from the European Centre for Medium-Range Weather Forecasts (ECMWF) \citep{hersbach2020era5}. This dataset provides a comprehensive view of Earth's atmospheric conditions, including detailed wind, temperature, and pressure fields. By analyzing these parameters, we aim to characterize a realistic decay profile for eddy-driven jets, potentially offering an analogy to those observed on Jupiter. We use monthly averaged zonal wind velocities spanning 1940-2024. Subsequently (Appendix~\ref{app:ERA5_processing}), we compute the zonally averaged zonal wind values (Figs.~\ref{fig:climatology},~\ref{fig:jet_per_pressure_level}) and track two distinct jet maxima in the southern hemisphere (Fig.~\ref{fig:earth_decay}). The southern hemisphere, being closer to an idealized aqua-planet, allows for the formation of an eddy-driven jet stream largely independent of the thermally-driven jet associated with the Hadley cell \citep{Vallis2006}.

We then compare the vertical structure of Earth's eddy-driven jet with the gravity inversion results and Jupiter GCMs (pink line and axis, Fig.~\ref{fig:GCMs}). Notably, the vertical structure features an Ekman (dissipative) layer at its base. This layer, caused by surface friction, results in a rapid wind decay. Above this, within the weather layer, a weak exponential decay is also apparent. While the vertical structure here bears some resemblance to solutions obtained for Jupiter via gravity inversion, no direct relationship between them is immediately apparent.

This comparative analysis of Jupiter GCMs and Earth's eddy-driven jets, when viewed alongside our gravity inversion results, underscores the inherent non-uniqueness in determining Jupiter's deep wind structure. While GCMs present a spectrum of decay profiles, from slowly decaying linear trends in deep models to rapid exponential decays in shallow models, and Earth's atmosphere offers a physically constrained example with distinct boundary layer effects, none provide a definitive, universally applicable solution. Each approach offers valuable insights, but the broad envelope of possible vertical structures revealed by gravity inversions, particularly with the introduction of stable layers, highlights that multiple physical mechanisms can produce similar observational signatures. Ultimately, integrating these diverse perspectives is crucial, yet further independent constraints are essential to resolve the degeneracy and precisely determine the true vertical extent and decay characteristics of Jupiter's jet streams.

 \section{Conclusions}\label{sec:Conclusions}

This study explored the profound implications of Jupiter's gravity field measurements from the Juno mission for understanding both its interior density structure and its atmospheric dynamics. We first established the inherent non-uniqueness in interpreting these gravity data, particularly concerning the interplay between the static density profile and the zonal wind structure. Our hierarchical approach then demonstrated the critical importance of including stable layers for accurately modeling Jupiter's gravity field and its jet streams vertical structure. Indeed, the plausible presence of SLs within Jupiter’s interior or atmosphere introduces a significant new dimension of degeneracy to this inverse problem.

Specifically, we demonstrated that shallow and extensive stable layers can profoundly alter the gravitational harmonics, necessitating a much faster decay of zonal winds with depth to maintain consistency with Juno’s observations. In contrast, deep stable layers were found to have negligible impact on the overall gravity field. This direct influence of stable layers is particularly profound in its ability to mimic the effects previously attributed to deeper wind penetration. {This degeneracy may have implications beyond Jupiter. For example, \citet{galanti2025observational} recently found that reproducing Saturn's Cassini gravity measurements requires zonal winds that are approximately $1.8$ times stronger than the observed cloud-level winds. While our study does not address Saturn directly, it demonstrates that stable layers can alter the gravity harmonics in a manner that partially mimics the effects of stronger or more deeply penetrating winds. Future investigations should therefore consider whether stable stratification introduces a similar degeneracy in the interpretation of Saturn's gravity field.} Our analysis further revealed complex decay patterns when simultaneously fitting both low-order and high-order odd harmonics, considerably expanding the envelope of physically permissible vertical wind structures. Notably, these findings are broadly consistent with theoretical predictions that convection-driven jet streams are expected to decay more rapidly in the presence of stable layers \citep{christensen2020,christensen2024quenching}. {However, when we explicitly impose jet decay within the stable layer, we generally fail to reproduce the observed Juno gravity harmonics. In particular, successful solutions require the zonal winds to retain at least $\sim10\%$ of their cloud-level amplitude down to depths of approximately $3000$\,km below the cloud tops ($\sim0.95\,R_J$). Therefore, while stable layers can facilitate jet attenuation, our results suggest that the primary decay region cannot coincide with the upper boundary of most proposed stable layers, and substantial wind amplitudes must persist below them.}

To contextualize these findings, we compared our gravity inversion solutions with results from two distinct Jupiter General Circulation Models, a deep convection-driven model and a shallow weather-layer model, as well as with ERA5 climatology data for Earth's eddy-driven jets. The GCMs presented a spectrum of decay profiles, ranging from slowly decaying linear-like trends to rapid exponential attenuation, none of which universally encompassed the full range of behaviors suggested by the gravity data, particularly when stable layers are considered. Similarly, while Earth's eddy-driven jets offered a valuable physical analog with its characteristic Ekman layer and exponential decay, direct extrapolation to Jupiter remains challenging due to vastly differing planetary conditions and the presence of unconfirmed mechanisms.

In conclusion, this work underscores that deciphering Jupiter's deep interior and the true nature of its powerful jet streams necessitates a holistic approach, explicitly accounting for the intricate coupling between the planet's density structure and its atmospheric dynamics. The expanded degeneracy introduced by plausible stable layers, combined with the varied predictions from GCMs and insights from Earth analogs, confirms that the envelope of physically permissible vertical wind profiles for Jupiter is considerably wider than previously assumed. This finding does not contradict previous studies that indicated the necessity of deep jets in a cylindrical projection to match the gravity data. However, our results emphasize that the precise vertical shape of the jet structure, and consequently, its underlying quenching mechanism, cannot be definitively established without further independent constraints. Future progress will therefore hinge upon acquiring new observational data or developing more sophisticated, self-consistent models that can robustly differentiate between these competing explanations and ultimately unveil the true nature of Jupiter's enigmatic interior.

 \section*{Data availability}
The TP$\rho$ profiles with stable layers will be available in a \textit{zenodo} repository upon publication.  The Juno gravity data is publicly available and the gravity coefficients presented here are taken from \citet{kaspi2023}. The ERA5 reanalysis data used in this study were obtained from the Copernicus Climate Change Service (C3S) Climate Data Store (CDS) \citep{hersbach2020era5}. Specifically, we utilized the "ERA5 monthly averaged data on pressure levels from 1940 to present" product. This dataset is publicly available and can be accessed free of charge at: https://cds.climate.copernicus.eu/datasets/reanalysis-era5-pressure-levels-monthly-means?tab=download.

\section*{Acknowledgements}
    This work is funded by the European Research Council (ERC) under the European Union’s Horizon 2020 research and innovation programme (grant agreement no. 101088557, N-GINE). K.~D.M. thank the Council for Higher Education in Israel for the CHE/PBC Fellowship for Postdoctoral training abroad for women for providing personal financial support. This publication is part of the project 'Exploring the Effects of Stable Stratified Layers on Jet Streams and Gravity Fields in Gas Giants' with file number 2024.045 of the research programme 'Computing Time on National Computer Facilities' which is (partly) financed by the Dutch Research Council (NWO) under the grant https://doi.org/10.61686/PHTUX64221. This publication is part of the project ENW.GO.001.001 of the research programme “Use of space infrastructure for Earth observation and planetary research (GO), 2022-1” which is (partly) financed by the Dutch Research Council (NWO).

\bibliographystyle{mnras}
\bibliography{kerensbib}

\begin{appendix}

\section{Comparison of the static density with published Jupiter profiles}
\label{app:compare}
\begin{figure}
   \centering
   \includegraphics[width=1\columnwidth]{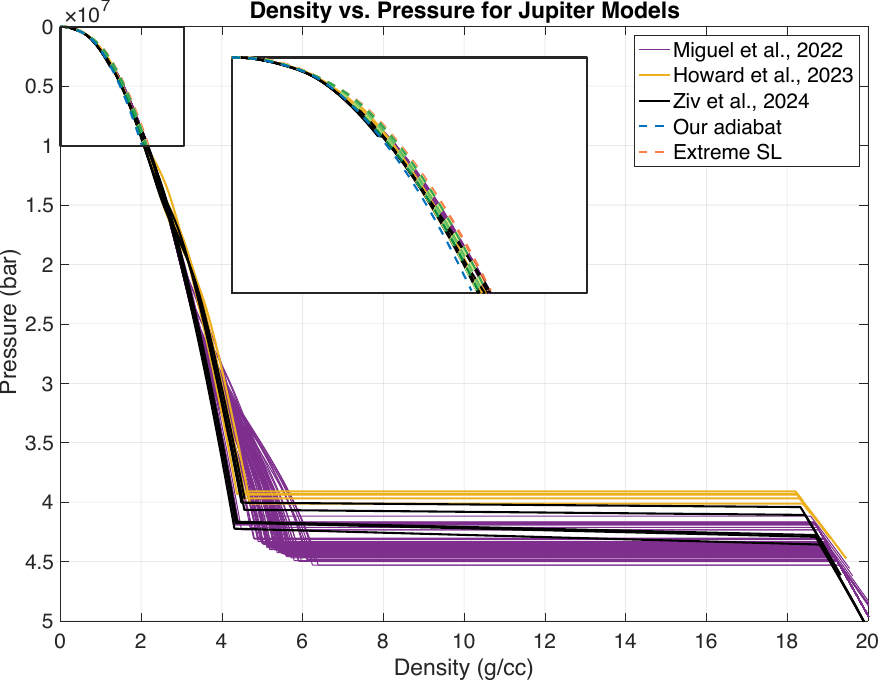}
   \caption{Density vs. Pressure for representative Jupiter models \citep{miguel2022,howard2023jupiter,ziv2024characterizing}, and the profiles from this work. In the legend we highlight the pure adiabat, and an Extreme SL ($1-10^5$ bar) case. Our profiles are deliberately shallow and do not include a core, illustrating that these atmospheric‑focused models are not intended to reproduce the low‑order even gravitational moments that depend on full interior structure.}
   \label{fig:Jupiter_profiles}
 \end{figure}

We compare density–pressure profiles from three published interior models \citep{miguel2022,howard2023jupiter,ziv2024characterizing} with our adiabatic profile and the most extreme stable-layer case (Fig.~\ref{fig:Jupiter_profiles}). From \citet{ziv2024characterizing}, we show the profiles closest to the averages of the four clusters identified in that work. The published models include deep interiors and a dilute core (and a compact core in some cases), with different emphases in each work. Our adiabat and Extreme SL are deliberately shallow, omit a core and interior composition adjustments, and thus illustrate how introducing shallow stable layers changes the outer-envelope density but cannot reproduce Jupiter’s mean density or the low-order even gravitational moments ($J_2$, $J_4$, $J_6$, $J_8$). Nonetheless, they are compatible with the atmospheric trends reported in these works and can be used to calculate the high-order and odd gravity harmonics, as explained in the main text.

     \begin{figure}
   \centering
   \includegraphics[width=1\columnwidth]{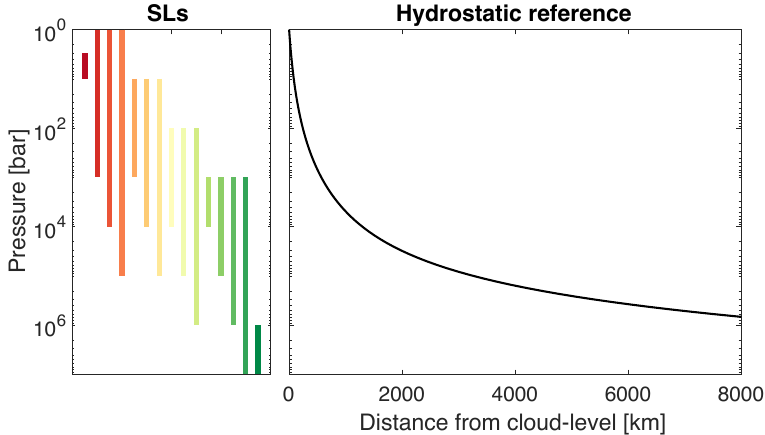}
   \caption{Pressure as a function of distance below the cloud tops, derived from the hydrostatic balance. The stable layers considered in this work are indicated in the left panel.}
   \label{fig:hydro}
 \end{figure}
 
\section{The optimization procedure}
\label{app:optim}

To determine the optimal vertical wind structure that best matches the observed gravitational field, we employ an inverse optimization approach following the methodology of \citet{Galanti2017b} and \citet{Kaspi2018}. The optimization is performed using the interior-point algorithm within MATLAB's \textit{fmincon} solver, which allows for the minimization of a non-linear cost function subject to constraints.

The objective of the optimization is to minimize the discrepancy between the measured and modeled gravitational harmonics. This is quantified by the cost function $L$, defined as:

\begin{equation}
L=\sum_{i\,}\sum_{j\,}w_{ij}\left(J_{i}{\rm ^{{\rm obs}}}-J_{i}{\rm ^{{\rm mod}}}\right)\left(J_{j}{\rm ^{{\rm obs}}}-J_{j}{\rm ^{{\rm mod}}}\right), \label{eq:cost}
\end{equation}

where $J_{n}{\rm ^{{\rm obs}}}$ represents the gravitational harmonics measured by Juno \citep{Iess2018,kaspi2023}, and $J_{n}{\rm ^{{\rm mod}}}$ denotes the harmonics calculated from the model given a specific wind structure and density profile. The term $w_{ij}$ is the weight matrix, which is the inverse of the covariance matrix of the measurements, effectively accounting for the measurement uncertainties and the correlations between different harmonics as detailed in \citet{Kaspi2018}. 

The indices $i$ and $j$ represent the specific gravitational harmonics included in the optimization process. We perform the optimization using two distinct sets of harmonics: the first considers only the low-order odd harmonics ($i,j=3,5,7,9$), while the second includes both the low-order odd harmonics and the high-order harmonics ($i,j=3,5,7,9 \dots 20$). By minimizing Eq.~\ref{eq:cost}, we identify the set of free parameters: either the decay coefficients in the constrained approach or the wind values at each grid point in the free-decay approach, that yields the statistically best-fitting vertical wind profile.

Note that, due to the inherent non-uniqueness of the inverse problem, the final optimal profile obtained in the free-decay approach exhibits a slight dependence on the initial conditions used for the optimization. Here, we initialize the wind strength at a dimensionless value of $0.9$ at all grid points to be optimized.
 
\section{GCM Data Processing for Jet Stream Analysis}
\label{app:GCM_processing}

This appendix details the methodology used to extract and process jet stream data from two distinct Jupiter GCMs. The aim was to obtain vertical profiles of zonal wind speeds for comparison with our gravity inversion results and Earth-based analogs. Both GCMs were selected for their ability to simulate eddy-driven jets, particularly relevant for Jupiter's mid-latitude dynamics.

\subsection*{1. Duer et al. (2023) Deep GCM}
\label{app:GCM_duer}
This GCM is a 'deep' model, driven by convecting plumes, developed using the Rayleigh convection code \citep{featherstone2016,Featherstone2022,duer2023}. This model simulates the deep interior dynamics of Jupiter.

\begin{figure}
   \centering
   \includegraphics[width=0.8\columnwidth]{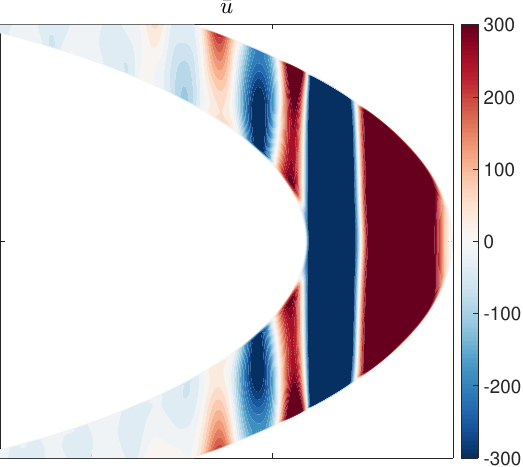}
   \caption{The zonally averaged zonal wind [cm s$^{-1}$] from the deep GCM by \citep{duer2023}. The figure illustrates a shell view and is showing latitudes $-45$ to $45$ in the outer shell.}
   \label{fig:duerGCM1}
 \end{figure}

\begin{figure}
   \centering
   \includegraphics[width=0.8\columnwidth]{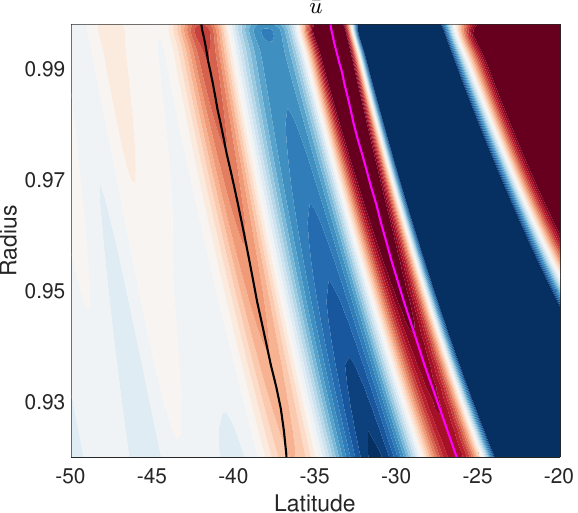}
   \caption{zoom in to the eddy driven jets in Fig.~\ref{fig:duerGCM1}. Black and pink lines are following to jet maxima in the southern hemisphere.}
   \label{fig:duerGCM2}
 \end{figure}

\begin{enumerate}
    \item {Data Extraction:} The 3D zonal velocity field (${u}$) was extracted from the time-dependent output of the GCM, regenerated by us with the control parameters and boundary conditions as specified in the original study.
    \item {Averaging:} The raw velocity field was first time-averaged to obtain a steady-state representation. Subsequently, longitudinal averaging was performed to derive zonally-averaged zonal wind profiles ($\overline{[u]}$) as a function of depth and latitude (Fig.~\ref{fig:duerGCM1}).
    \item {Jet Identification:} Within the Southern Hemisphere, two distinct jet maxima were identified and tracked vertically (Fig.~\ref{fig:duerGCM2}). These represent the core velocities of specific eddy-driven jets. While this GCM exhibits near-perfect north-south symmetry, the Southern Hemisphere was chosen for consistency with the ERA5 analysis.
    \item {Normalization:} To facilitate comparison with other models and the gravity inversion results, the extracted vertical profiles of the jet maxima were normalized by their respective maximum speeds.
    \item {Output:} The processed vertical profiles are shown in Figure~\ref{fig:GCMs} in the main text. This model, lacking an explicit internal friction mechanism, typically exhibits a relatively slow, linear decay of jet strength with depth.
\end{enumerate}

\subsection*{2. Young et al. (2019) Shallow GCM}
\label{app:GCM_young}
The second GCM is a 'shallow' model from \citet{young2019}, implemented using the MITgcm \citep{Adcroft2007}. This model focuses on Jupiter's weather layer, solving for atmospheric dynamics between approximately 0.5 and 18 bars, driven by solar forcing and parameterized internal heat flux.

\begin{figure}
   \centering
   \includegraphics[width=0.9\columnwidth]{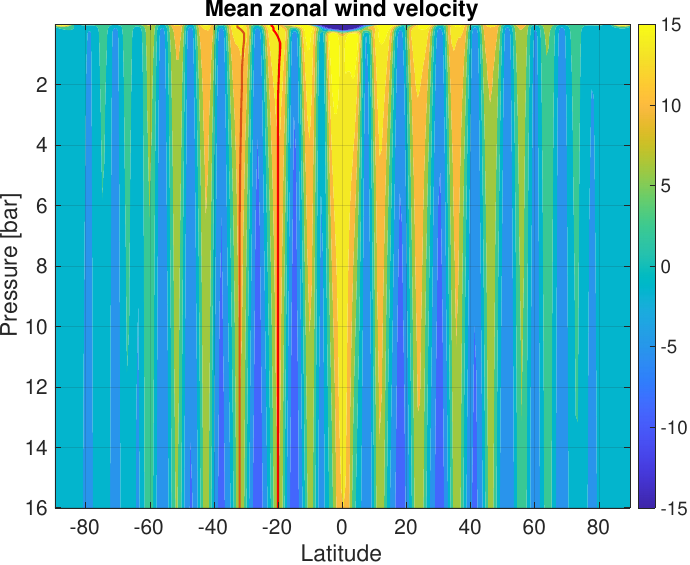}
   \caption{The zonally averaged zonal wind [m s$^{-1}$] from the shallow GCM by \citep{young2019}. Orange and red lines are following two jet maxima in the southern hemisphere.}
   \label{fig:youngGCM}
 \end{figure}

\begin{enumerate}
    \item {Data Extraction:} We utilize 'run B' from \citet{young2019,young2019b}, which includes an interior heat flux of $5.7$ W m$^{-2}$. The 3D zonal velocity field (${u}$) was extracted from the output variables provided at: https://ora.ox.ac.uk/objects/uuid:ce0e862f-2cf4-400e-b558-60889e316eda.
        \item {Averaging:} Similar to the deep GCM, the velocity field was time-averaged (6 snapshots are provided) and then longitudinally averaged to obtain zonally-averaged zonal wind profiles ($\overline{[u]}$) as a function of depth and latitude. The original study by \citet{young2019} provides a detailed analysis of the force balance in this model, confirming the dominance of eddy terms in mid-latitudes.
    \item {Jet Identification:} Two distinct jet maxima in the Southern Hemisphere were identified and tracked vertically through the simulated weather layer (Fig.~\ref{fig:youngGCM}).
    \item {Normalization:} The vertical profiles of the jet maxima were normalized by their respective maximum speeds for comparative analysis.
    \item {Output:} The processed vertical profiles are included in Figure~\ref{fig:GCMs} in the main text. This shallow model typically demonstrates a rapid, exponential decay of wind strength with depth, particularly within the upper part of the weather layer, before stabilizing at greater depths within its domain.
\end{enumerate}

\section{ERA5 Data Processing Methodology}\label{app:ERA5_processing}

The ERA5 reanalysis dataset, provided by the European Centre for Medium-Range Weather Forecasts (ECMWF) \citep{hersbach2020era5}, offers a comprehensive view of Earth's atmospheric conditions with high temporal and spatial resolution. We extracted monthly averaged zonal wind velocity (variable 'u') data from the ERA5 reanalysis for the period spanning 1940 to 2024. This extensive temporal coverage allows for the analysis of robust climatological trends. Our primary focus was on the Southern Hemisphere, as its atmospheric dynamics are often considered a closer analog to an idealized aqua-planet, facilitating a clearer examination of eddy-driven jet stream formation independent of thermally-driven systems like the Hadley cell \citep{Vallis2006}.

\begin{figure}
   \centering
   \includegraphics[width=0.8\columnwidth]{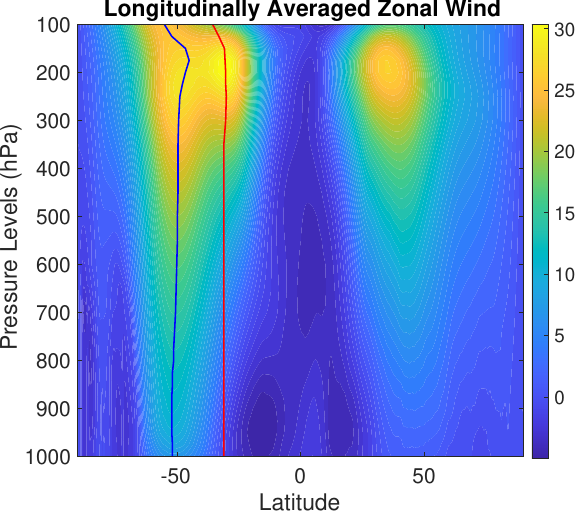}
    \caption{Climatological mean longitudinally averaged zonal wind (m/s) from ERA5 reanalysis (1940-2024). Color contours show wind speed as a function of latitude and pressure. The blue line traces the core of the eddy-driven jet, and the red line traces the core of the thermally-driven jet, both in the Southern Hemisphere. The $1000$ hpa level is the Earth's surface.}
   \label{fig:climatology}
 \end{figure}

\begin{figure}
   \centering
   \includegraphics[width=\columnwidth]{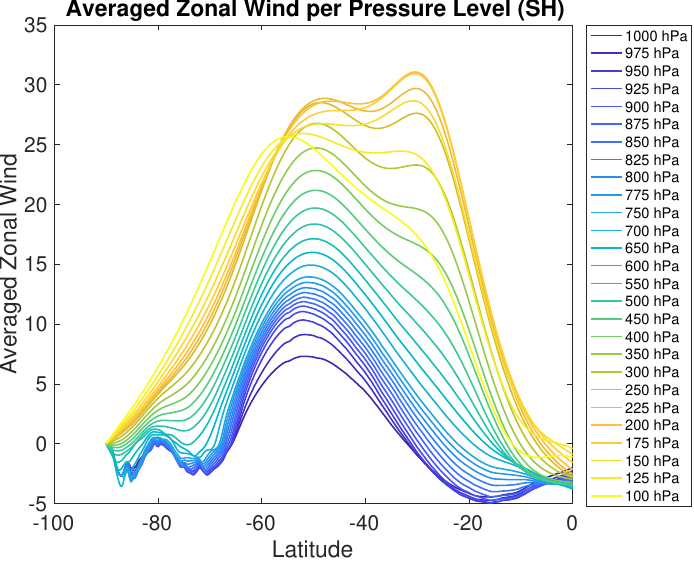}
    \caption{Climatological mean zonal wind velocity (m/s) as a function of latitude, shown for selected pressure levels in the Southern Hemisphere (ERA5 data, 1940-2024). Each colored line corresponds to a specific pressure level, identified in the legend. This illustrates the the thermally-driven jet vanishes at around $500$ hpa, and below the eddy-driven jet dominates.}
   \label{fig:jet_per_pressure_level}
 \end{figure}

\begin{figure}
   \centering
   \includegraphics[width=0.8\columnwidth]{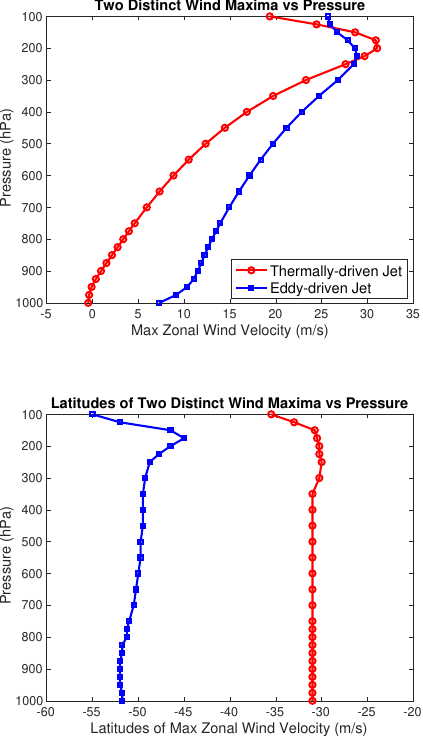}
    \caption{Vertical profiles of the two dominant Southern Hemisphere jet streams derived from ERA5 climatology (1940-2024). (top) Maximum Zonal Wind Velocity: Vertical decay of the thermally-driven jet (red line) and the eddy-driven jet (blue line) as a function of pressure.  (bottom) Latitudes of Maximum Zonal Wind Velocity: Evolution of the core latitude for each jet with pressure.}
   \label{fig:earth_decay}
 \end{figure}

To identify and track the dominant eddy-driven jets in the Southern Hemisphere, the following procedure was applied to the climatological zonal wind field:
\begin{enumerate}
    \item {Zonal and Temporal Averaging:} For each month in the selected period, the zonal wind ('u') component was first averaged over all longitudes to obtain monthly zonally-averaged zonal winds. These monthly averages were then accumulated and averaged over the entire 1940-2024 time span to compute a long-term climatological mean zonal wind field (Fig.~\ref{fig:climatology}).
    \item {Jet Identification per Pressure Level:} For each selected pressure level, the zonally-averaged zonal wind profile within two distinct Southern Hemisphere latitude bands was analyzed. A peak-finding algorithm was employed to identify the primary local maximum in wind speed within each band, corresponding to the core of an eddy-driven jet stream  (Fig.~\ref{fig:jet_per_pressure_level}).
    \item {Manual Refinement:} In instances where the automated peak-finding algorithm was ambiguous or failed to identify a consistent jet maximum across all pressure levels (e.g., near the bottom of the atmospheric column or in regions of weaker flow), manual adjustments were made. This involved selecting the most probable jet core latitude and corresponding wind speed based on visual inspection of the broader wind field to ensure continuity in the tracked jet's vertical profile (Fig.~\ref{fig:earth_decay}).
\end{enumerate}

\end{appendix}

\bsp	
\label{lastpage}
\end{document}